\documentclass[preprint]{aastex631}

\usepackage{hyperref}
\usepackage{amsmath}
\usepackage{enumerate}
\usepackage[utf8]{inputenc}
\usepackage{cleveref}
\usepackage{amssymb}
\usepackage[T1]{fontenc}
\usepackage{enumitem}
\usepackage{url}
\usepackage[misc]{ifsym}
\usepackage{multirow}

\begin{document}

\title{Polarization Properties of the Electromagnetic Response to High-frequency Gravitational Wave}

\author[0000-0002-1190-473X]{Jian-Kang Li}
\affiliation{Institute for Frontiers in Astronomy and Astrophysics, Beijing Normal University, Beijing 102206, China}
 \affiliation{School of Physics and Astronomy, Beijing Normal University, Beijing 100875, China.}%Lines break automatically or can be forced with \\
 \author[0000-0001-7906-0919]{Wei Hong}
 \affiliation{Institute for Frontiers in Astronomy and Astrophysics, Beijing Normal University, Beijing 102206, China}
  \affiliation{School of Physics and Astronomy, Beijing Normal University, Beijing 100875, China.}
\author[0000-0002-3363-9965]{Tong-Jie Zhang \href{mailto:tjzhang@bnu.edu.cn}{\textrm{\Letter}}}
 \affiliation{Institute for Frontiers in Astronomy and Astrophysics, Beijing Normal University, Beijing 102206, China}
 \affiliation{School of Physics and Astronomy, Beijing Normal University, Beijing 100875, China.}



\begin{abstract}

Electromagnetic waves (EMWs) can be generated by gravitational waves (GWs) within a magnetic field via the Gertsenshtein effect. The conversion probability between GWs and EMWs can be enhanced by inhomogeneities in the electron density and magnetic field within the magnetized plasma of both the Milky Way (MW) and the intergalactic medium in the expanding Universe. Polarized GWs can induce polarized EMWs, and the polarization properties of these EMWs can be altered by Faraday rotation as they propagate through magnetized plasma. Additionally, the polarization intensity of the EMWs may be weakened due to depolarization effects. In this study, we calculate the enhanced GW-EMW conversion in inhomogeneous magnetized plasma during the propagation of GWs through the Universe and our galaxy. We analyze the polarization states of the EMWs generated by polarized GWs and discuss the depolarization effects induced by the medium's irregularities, as well as the differential Faraday rotation occurring in multilayer polarized radiation. Our work provides alternative methods for detecting GWs and exploring their polarization states, and potentially constraining the parameters of the possible GW sources, especially the primordial black hole, contributing to the advancement of very-high-frequency GW detection and research.

\end{abstract}

\keywords{\href{http://asothetrsaurus.org/uat/678}{Gravitational waves(678)}; \href{http://astrothesaurus.org/uat/1278}{Polarimetry(1278)};\href{http://astrothesaurus.org/uat/1338}{Radio astronomy(1338)}}


\section{\label{sec:introduction}Introduction}
Several GW events have been reported by LIGO and Virgo 
\citep{PhysRevLett.116.061102,PhysRevLett.116.241103,PhysRevLett.118.221101,2017ApJ...848L..12A,PhysRevLett.119.141101,PhysRevLett.119.161101,PhysRevX.6.041015}, indicating the opening of a new window in astronomy and astrophysics and offering unprecedented opportunities to explore the universe. Ground-based detectors like LIGO, Virgo, and KAGRA are highly sensitive to GWs in the frequency range of a few Hz to several kHz \citep{PhysRevD.93.112004,2015CQGra..32b4001A,2012CQGra..29l4007S}, which typically originate from events such as stellar-mass binary black hole or neutron star mergers. Recently, the stochastic gravitational wave background in the nHz range, has been detected by pulsar timing arrays (PTAs) \citep{2023ApJ...951L...8A,2023A&A...678A..50E,2023ApJ...951L...6R,2023RAA....23g5024X}. Looking ahead, space-based detectors such as  LISA and its variants (e.g., eLISA) are expected to probe the mHz frequency band \citep{2017arXiv170200786A,2012CQGra..29l4016A}, enabling the detection of signals from sources like massive black hole mergers \citep{2013GWN.....6....4A,PhysRevD.93.024003}, galactic binaries \citep{2014LRR....17....3P}, and potentially primordial gravitational waves \citep{2016JCAP...04..001C}, thus bridging the gap between ground-based detectors and PTAs. 

GWs with frequencies above 10 kHz have so far received relatively little attention in both theoretical research and instrument development, primarily due to the challenges associated with their detection and the limited sensitivity of current detectors in this frequency range. On the other hand, these very-high-frequency gravitational waves (VHFGWs) are anticipated to originate from processes in the early universe \citep{Fomin19,2022GReGr..54..156C}. Potential origins include some types of cosmic inflation \citep{2005JCAP...01..005K,2010Sci...328..989K,2015JCAP...06..040P,2020JCAP...09..015E}, pre-big-bang \citep{2003PhR...373....1G,2016JCAP...12..010G}, Kaluza-Klein gravitons from braneworld models \citep{2007CQGra..24F..33C,PhysRevD.88.064005,PhysRevD.97.104053}, and primordial black holes \citep{PhysRevD.84.024028,PhysRevD.89.103501,PhysRevD.104.023524,2023JCAP...02..062G}. Detecting and studying these VHFGWs could provide critical insights into the dynamics of the early universe and reveal phenomena beyond the reach of conventional astrophysical and cosmological observations, probing energy scales and phenomena beyond the reach of current particle accelerators. 

Despite the challenges in detecting VHFGWs, primarily owing to their short wavelengths requirement for the detector components of comparable scales and complicating the design and implementation of effective detection systems, several innovative methods have been proposed to enable the detection of VHFGWs, encompassing both experimental setups and observational techniques. One approach involves magnon-based detectors, which exploit the interaction between gravitational waves and magnons—quanta of spin waves in magnetic materials, where passing VHFGWs can induce resonant excitations, producing detectable signals \citep{2020EPJC...80..179I,2020EPJC...80..545I,2023EPJC...83..766I}. Another approach involves detecting subtle frequency shifts in laser beams induced by VHFGWs, using techniques like optical demodulation or atomic clocks for precise measurements \citep{PhysRevD.108.L061303}. Additionally, levitated sensor detectors involve optically levitated dielectric particles acting as sensitive resonant sensors for GWs, with systems tunable to specific frequency ranges, offering potential detection capabilities in the 50-300 kHz \citep{PhysRevLett.110.071105}. Furthermore, GWs conversion into EMWs via the Gertsenshtein effect \citep{gertsenshtein1962wave} has also been extensively studied under controlled laboratory settings \citep{2021JCAP...03..054R,2021LRR....24....4A,PhysRevD.105.116011,PhysRevLett.129.041101,2022Symm...14.2165T}, as well as theoretically predicted for potential detection through astronomical observation methods on astrophysical or cosmic scales \citep{1994PhLB..336..431D,PhysRevD.68.044017,PhysRevD.80.042002,2012JCAP...12..003D,PhysRevD.87.124029,PhysRevD.99.044022,2020EPJC...80..879L,PhysRevD.102.103501,PhysRevLett.126.021104,2024PTEP.2024b3E03I,PhysRevD.109.063026,PhysRevLett.132.131402,2024PhLB..85138574A,2024JCAP...05..051H,PhysRevD.110.083042,PhysRevD.110.103003}.

Radio telescopes equipped with modern dual-polarization receivers can comprehensively measure the polarization state of interstellar EMWs \citep{2014JKAS...47...15T,2021hai1.book..127R}. In this work, we focus mainly on the GW-EMW conversion within the radio band. Several works have proposed to probe GWs using radio telescopes to search EMWs produced from the Gertsenshtein effect occurring in the early Universe \citep{PhysRevD.102.103501,PhysRevLett.126.021104,2024PhLB..85138574A,2024JCAP...05..051H}, in the Milky Way (MW) magnetic field \citep{PhysRevD.110.083042}, in the planetary magnetic field \citep{2024PTEP.2024b3E03I,PhysRevLett.132.131402}, or in the magnetic field of neutron stars \citep{PhysRevD.109.063026,2024arXiv241205338H}. Nevertheless, the polarization features of the EMWs produced from the Gertsenshtein effect have seldom been discussed. When polarized EMWs propagate through magnetized plasma, phenomena such as Faraday rotation \citep{1953JAP....24..604K}, Faraday depolarization \citep{1966MNRAS.133...67B}, Cotton-Mouton Effect \citep{cotton1905phenomene,cotton1905birefringence}, cyclotron resonance and plasma oscillation may occur. These effects can modify the polarization states, intensity, and propagation characteristics of EMWs, thereby providing critical insights into the plasma properties and magnetic field configurations. Magnetic fields and plasma are believed to be ubiquitously distributed across the universe \citep{1985IAUS..107..537K,1994RPPh...57..325K,2001PhR...348..163G,2015bps..book.....C}, spanning hierarchical scales from our Galaxy to galaxy clusters and even the cosmic web, where they exhibit structural perturbations at different magnitudes \citep{2005ApJ...629..139S,2013A&ARv..21...62D,2020PPCF...62a4014F,2023ChA&A..47..490W}. 

Fluctuations in plasma density and the magnetic field can enhance the GW-EMW conversion probability during the propagation, consequently leading the photon flux spatial accumulation in the propagating direction. Specifically, we aim to analyze the polarization properties of EMWs produced by the Gertsenshtein effect in the high-redshift universe. For an observer on Earth, these EMWs would traverse magnetized plasma in the cosmic web and MW before reaching radio telescopes, thereby manifesting distinct polarization signatures.

The structure of this work is organized as follows.  In Section \ref{sec:GW-EMWConversion}, we present the calculations of the GW-EMW conversion across MW, the IGM in the larger universe, taking into account perturbations in electron densities, magnetic fields, and the effects of cosmic expansion. In Section \ref{sec:Polarization Properties}, we present the polarization properties of the EMWs in radio observations. Section \ref{sec:Sensitivity} provides predicted constraints derived from currently operating and future radio observational facilities. Discussion and conclusion are made in Section \ref{sec:DiscussionandConclusion}.

Unless otherwise stated, all calculations are performed using MKS units, and the metric signature $\eta = \mathrm{diag}(-+++)$ is adopted. The standard $\lambda$CDM cosmology parameters we adopt in this work are \citep{2009ApJ...707..916F,2020A&A...641A...6P}: matter density parameter $\Omega_m=0.315$, dark energy density parameter $\Omega_{\Lambda}=0.6847$, Hubble constant $H_0=100h_o=67.4 \, \mathrm{km\cdot s^{-1}\cdot Mpc^{-1}}$ with $h_0$ being a dimensionless factor, and blackbody CMB temperature $T_\mathrm{CMB}=2.725 \mathrm{K}$.

\section{Gravitational wave-Electromagnetic wave Conversion}\label{sec:GW-EMWConversion}

Generally, we start with the total action of the GW-EMW system
\begin{equation}
  \mathcal{S} =\mathcal{S}_\mathrm{GW}+\mathcal{S}_\mathrm{EM},
  \label{totaction}
\end{equation}
where $\mathcal{S}_\mathrm{GW}$ and $\mathcal{S}_\mathrm{EM}$ are the action of gravitational wave and electromagnetic wave, respectively. The two terms are given by
\begin{eqnarray}
  \mathcal{S}_\mathrm{GW}&&=\frac{1}{\kappa^2}\int{d^4x\sqrt{-g}R} ,\nonumber\\
  \mathcal{S}_\mathrm{EM}&&=-\int{d^4x\sqrt{-g} \left(\frac{1}{4\mu_0}F^{\mu\nu}F_{\mu\nu} - A_{\mu} J^{\mu} \right)   },
  \label{actionGWEM}
\end{eqnarray}
where $R$ is the Ricci scalar, $g$ is the metric determinant, $\kappa^2=16\pi G_N/c^4$ with $G_N$ being the Newtonian constant, The electromagnetic field tensor is defined as $F_{\mu\nu} = \partial_\mu A_\nu - \partial_\nu A_\mu$, where $A_\mu$ is the vector potential, $\mu_0$ is the vacuum permeability, and $J^\mu$ represents the current density. For GWs, the metric can be expended as small small perturbation $h_{\mu\nu}$ to the at Minkowski spacetime $\eta_{\mu\nu}$, i.e.,
\begin{equation}
  g_{\mu\nu} =\eta_{\mu\nu}+h_{\mu\nu}.
  \label{GWmetric}
\end{equation}
In the presence of a weak magnetic field, where the plasma frequency exceeds the cyclotron frequency, the current density in the magnetized plasma can be approximated as \citep{fitzpatrick2022plasma}:
\begin{equation}
  J^\mu=-\epsilon_0 \omega_{pl}^2 A^\mu,
  \label{currentdensity}
\end{equation}
where $\omega_{pl}=\sqrt{n_e e^2/\epsilon_0 m_e}$ is the electron plasma frequency with $n_e$, $e$, $\epsilon_0$ and $m_e$ being the electron number density, the electron charge, vacuum permittivity and the electron mass, respectively. Considering waves propagating in magnetized plasma with external magnetic field $\boldsymbol{B}=(B_x, B_y, B_z)$ along the $z$-direction, $\partial_i=(0, 0, \partial_z)$, and we apply Lorenz gauge for EM and transverse-traceless (TT) for GW, while neglecting higher-order terms, from Eq. (\ref{totaction})-(\ref{currentdensity}) we can obtain the following equations \citep{2012JCAP...12..003D,PhysRevLett.126.021104,2024JCAP...05..051H}
\begin{subequations}
  \label{MaxwellEinsteineq}
  \begin{eqnarray}
    \left(\square-\frac{\omega_{pl}^2}{c^2}\right)  \begin{pmatrix} A_x  \\ A_y \end{pmatrix}= \begin{pmatrix}  B_y\partial_z h_{+}-B_x\partial_z h_{\times}  \\  B_y\partial_z h_{\times}+B_x\partial_z h_{+} \end{pmatrix},\label{Maxwelleq} \\ 
    \square  \begin{pmatrix}  h_{+}  \\  h_{\times} \end{pmatrix}=\kappa^2 \begin{pmatrix} -B_x\partial_z A_y -B_y\partial_z A_x   \\ B_x\partial_z A_x -B_y\partial_z A_y \end{pmatrix},
    \label{Einsteineq}
  \end{eqnarray}
\end{subequations}
where $\square=\partial_t^2/c^2-\partial_z^2$ is the d'Alembert operator, $h_{+}$ and $h_{\times}$ are the GW polarizations in the TT gauge. The fields $A_i$ and $h_{ij}$ can be expended into Fourier modes
\begin{eqnarray}
  A_i (z,t)=&&\sum_{\lambda=x,y} \int_{-\infty}^{+\infty}A_\lambda (z, \omega) e^{-i\omega t},\nonumber\\
  h_{ij}(z,t)=&&\sum_{\lambda=+,\times} \int_{-\infty}^{+\infty}h_\lambda (z, \omega) e^{-i\omega t}.
  \label{Fouriereq}
\end{eqnarray}
Eq. (\ref{MaxwellEinsteineq})-(\ref{Fouriereq}) can be linearized by the slowly varying envelope approximation (SVEA), then we can obtain first-order differential equations \citep{2018EPJC...78...63E,2019EPJC...79..231E,2019EPJC...79.1032E,2024JCAP...05..051H}
\begin{equation}
  \left(\frac{\omega}{c}+i\partial_z+\mathcal{M}\right)\Psi\left(z,\omega\right)=0,
  \label{lineardiffeqEMGW}
\end{equation}
where $\Psi(z,\omega)=(h_\times/\kappa,h_+/\kappa,\sqrt{n}A_x,\sqrt{n}A_y)^{\mathrm{T}}$ is the four component field with $n=\sqrt{1-\omega_{pl}^2/\omega^2}$ being the refraction index, and $\mathcal{M}(z,\omega)$ is the mixing matrix determined by the dispersion relation of GWs and EMWs in the magnetized plasma. 

For monochromatic wave, the dispersion relation may be affected by plasma oscillation, cyclotron resonance, Cotton-Mouton effect and QED effect. For typical electron density $n_e\lesssim 10^{6} \ \mathrm{m^{-3}}$ and magnetic filed $B\lesssim 10^{-10} \ \mathrm{T}$ in MW and galaxy cluster, plasma frequency dominates over the cyclotron frequency $\omega_c=eB/m_e$. Additionally, both the Cotton-Mouton and QED effects are proportional to the square of the transverse magnetic field component $B_\perp^2$, making their contributions relatively small in such environments. Consequently the dispersion is primarily governed by the plasma frequency \citep{2012JCAP...12..003D,PhysRevD.87.124029,PhysRevD.99.044022,2019EPJC...79..231E,2024PhLB..85138574A,2024JCAP...05..051H}. From Eq. (\ref{MaxwellEinsteineq})-(\ref{lineardiffeqEMGW}), we can obtain the mixing matrix
\begin{equation}
  \mathcal{M}=\begin{pmatrix} 
  0 & 0 & \dfrac{-i\kappa n B_{x}}{2} & \dfrac{i\kappa n B_{y}}{2}\\ 
  0 & 0 & \dfrac{i\kappa n B_{y}}{2} & \dfrac{i\kappa n B_{x}}{2} \\
  \dfrac{i\kappa  B_{x}}{n+1} & \dfrac{-i\kappa  B_{y}}{n+1} & \dfrac{\omega}{c}(n-1) &0 \\
  \dfrac{-i\kappa  B_{y}}{n+1} & -\dfrac{i\kappa  B_{x}}{n+1} & 0 & \dfrac{\omega}{c}(n-1)
  \end{pmatrix}.
\end{equation}

The initial solutions of EMWs should be $A_x(z,\omega)=A_y(0,\omega)=0$. Therefore, the EMWs solutions can be written as 
\begin{eqnarray}
  A_x(z,\omega)&&=\dfrac{\sqrt{n} l_{o}}{n+1} \sin\left(\dfrac{z}{l_{o}}\right) e^{ik_\gamma z}B_y h_+(0,\omega)\nonumber\\
  &&-\dfrac{\sqrt{n} l_{o}}{n+1} \sin\left(\dfrac{z}{l_{o}}\right) e^{ik_\gamma z}B_xh_\times(0,\omega),\nonumber\\
  A_y(z,\omega)&&=\dfrac{\sqrt{n} l_{o}}{n+1} \sin\left(\dfrac{z}{l_{o}}\right) e^{ik_\gamma z}B_y h_\times(0,\omega)\nonumber\\
  &&+\dfrac{\sqrt{n}\ l_{o}}{n+1} \sin\left(\dfrac{z}{l_{o}}\right) e^{ik_\gamma z}B_xh_+(0,\omega)
  \label{EMWsolution}
\end{eqnarray}
where $h_\times(0,\omega)$ and $h_+(0,\omega)$ are the initial amplitudes of GWs at $z=0$, and $l_{o}=2/(k_\gamma-k_g)$ is oscillation length defined by
\begin{equation}
  l_{o}=\frac{2}{\sqrt{(1-n)^2\omega^2/c^2+\kappa^2(B_x^2+B_y^2)}}.
\end{equation}
Then the GW-EMW conversion probability after traversing a distance $\Delta L=L-L_0$ can be calculated by
\begin{eqnarray}
  \mathcal{P}(\Delta L)&&=\dfrac{\langle \vert A_x(\Delta L)\vert ^2\rangle+ \langle \vert A_y(\Delta L)\vert ^2\rangle}{\langle \vert h_+(L_0)\vert ^2\rangle+ \langle \vert h_\times(L_0)\vert ^2\rangle}\nonumber\\
  &&=\dfrac{n\kappa^2(B_x^2+B_y^2)}{(n+1)^2}l_{o}^2\sin^2{\left(\dfrac{\Delta L}{l_{o}}\right)}
  \label{conversionprob}
\end{eqnarray}
For small phase difference between GW and EMW, i.e. $\Delta L\ll l_{o}$, the magnetic field and plasma density still stay highly homogeneous,and Eq. (\ref{conversionprob}) can be approximated into
\begin{equation}
  \mathcal{P}(\Delta L)\approx\dfrac{n\kappa^2(B_x^2+B_y^2)}{(n+1)^2}(\Delta L)^2.
  \label{conversionprobsamllapprox}
\end{equation}
While the typical case is $\Delta L\gg l_{osc}$ and the probability can be averaged as
\begin{equation}
  \overline{\mathcal{P}} \approx\dfrac{n\kappa^2(B_x^2+B_y^2)}{2(n+1)^2}l_{o}^2.
  \label{conversionproblargeapprox}
\end{equation}
When fluctuations of magnetic field and plasma density are taken into consideration, the phase difference between GWs and EMWs can shift by $\pi$due to variations in the plasma refractive index or magnetic field gradients. This enables quasi-phase matching, where the phase mismatch between GWs and EMWs is quasi-periodically compensated. As a result, the energy of the EMWs can be accumulated. The electron density and magnetic filed can be expressed as the sum of static term and perturbation term with zero mean, given by 
\begin{subequations}
  \begin{eqnarray}
  n_e(\boldsymbol{r})&&=n_{e0}+\tilde{n}_e(\boldsymbol{r})\\\label{neperturbation}
  \boldsymbol{B}(\boldsymbol{r})&&=\boldsymbol{B}_{0}+\tilde{\boldsymbol{B}}(\boldsymbol{r})\label{Bperturation}
  \end{eqnarray}
\end{subequations}
where $\langle \tilde{n}_e(\boldsymbol{r})\rangle =0$ and $\langle \tilde{\boldsymbol{B}}(\boldsymbol{r})\rangle =0$. Since the typical electron density and magnetic field we discuss are relatively tiny in order-of-magnitude compared with the radio frequencies $\omega\gtrsim 10^6 \ \mathrm{Hz}$, the variations of $\omega_{pl}^2$ term and $\kappa (B_x^2+B_y^2)$ term can be neglected in the expression of $l_{osc}$, Additionally, the $n/(n+1)^2$ term can be approximated into 1/4. Assuming that that spatial fluctuations of the magnetic field are statistically homogeneous and isotropic, the magnetic field can be written as its Fourier components
\begin{equation}
  \tilde {\boldsymbol{B}}(\boldsymbol{q}_B)=\int d^3r \tilde{\boldsymbol{B}}(\boldsymbol{r})e^{i\boldsymbol{q}_B\cdot \boldsymbol{r}}.
\end{equation}
The second moment of magnetic field is \citep{2013A&ARv..21...62D}
\begin{eqnarray}
\langle\boldsymbol{B} (\boldsymbol{r})\boldsymbol{B}^*(\boldsymbol{r})\rangle =&& \int d^3q d^3q'\langle \boldsymbol{B} (\boldsymbol{q}_B) \boldsymbol{B}^*(\boldsymbol{q'}_B) \rangle e^{i(\boldsymbol{q_B}-\boldsymbol{q'}_B)\cdot\boldsymbol{r}}
  \nonumber\\
  =&& \int d^3q P_B(q_B),
\end{eqnarray}
where $P_B(q_B)$ is the magnetic field spectrum. For $q_B\ge2l_{o}$, i.e. $s_B\le \pi l_{o}$ where $s_B=2\pi/q_B$, the GW-EMW stay quasi phase matching in $\Delta L/s_B$ independent regions with a conversion probability $\mathcal{P}(\Delta L)$ each (The detailed expressions of conversion probability in different case and spectrum models are given by \cite{2024PhLB..85138574A}), and the total conversion probability is \citep{PhysRevD.54.4757,PhysRevD.80.042002,PhysRevLett.126.021104,2024PhLB..85138574A} 
\begin{equation}
  \mathcal{P}(\Delta L)\approx\dfrac{n\kappa^2(B_x^2+B_y^2)}{2(n+1)^2}l_{o} \Delta L,
  \label{conversionprobquasiapprox}
\end{equation}
where 
\begin{equation}
  B_x^2+B_y^2=B_{0x}^2+B_{0y}^2+\frac{2}{3}\int d^3 qP_B(q_B).
\end{equation}
Theoretically, the power spectrum of magnetic field perturbation is usually modeled by power-law spectrum \citep{2005PhR...417....1B,2011PhRvE..83f5401R}
\begin{equation}
  P_B(q_B)=C_B^2(s)q_B^{-\alpha_B},
  \label{powerlawB}
\end{equation} 
where $C_B^2(s)$ is the coefficient of the magnetic field perturbation power-law spectrum and $\alpha_B$ is the corresponding power index. Observationally, power-law spectra are also present in different astrophysical scales (e.g. \cite{2004ApJ...610..820H,2006Sci...311..827I,2011A&A...529A..13K,2017A&A...603A.122G,2018CQGra..35o4001H,2024A&A...683A.114K}). 

For GWs produced in the early age of the universe, the redshift should be taken into account. The electron number density at redshift $z_s$ is $n_e(z_s)=n_{b0}(1+z_s)^3X_e(z_s)$ where $n_{b0}$ is the baryon number density today measured by \cite{2020A&A...641A...6P} and $X_e(z_s)$ is the ionization fraction simulated by \cite{2015JCAP...12..028K}. The magnetic field is related to the electron number density by $B(z_s)=B_0(n_e/n_{b0})^{2/3}$, where $B_0$ is the co-moving cosmological magnetic field at $z_s=0$. The co-moving cosmological magnetic field can also modeled by $B_0(z_s)=B_0(1+z_s)^{-\alpha}$ due to some unspecified magnetic field injection, amplification or evolution mechanism \citep{2022MNRAS.515..256P}. The refraction index then can be approximated to $n(z_s)\approx1-{n_{b0}(1+z_s)X_e(z_s)e^2}/{(2\epsilon m_e \omega_0^2)}$, where $\omega_0$ is the observation frequency. The oscillation length, considering the domination of plasma frequency term, can be approximated as $l_{o}\approx4\epsilon_0m_e\omega_0c/[n_{b0}(1+z_s)^2X_e(z_s)e^2]$. Then the conversion probability Eq. (\ref{conversionprobquasiapprox}) can be rewritten into
\begin{eqnarray}
  \mathcal{P}(z_s)&&=\int_{0}^{z_s}\dfrac{n(z_s')\kappa^2(B_x^2(z_s')+B_y^2(z_s'))}{2[n(z_s')+1]^2}l_{o}(z_s') dl(z_s')\nonumber\\
  &&\approx\dfrac{(B_{0x}^2+B_{0y}^2)\kappa^2\epsilon_0 m_e \omega_0c^2}{n_{b0}e^2H_0}\int_{0}^{z_s}(1+z_s')^{1-2\alpha}X_e^{1/3}\nonumber\\
  &&\times \dfrac{d z_s'}{\sqrt{\Omega_m(1+z_s')^3+\Omega_{\Lambda}}},
\end{eqnarray}
where $dl(z_s)$ is the proper displacement at redshift $z_s$ given by \citep{1999ApJ...514L..79B}
\begin{equation}
  dl(z_s)=\frac{c \, dz_s}{H_0 (1 + z_s) \sqrt{\Omega_{m} (1 + z_s)^3 + \Omega_{\Lambda}}}.
  \label{properdistance}
\end{equation}

\section{Polarization Properties of Electromagnetic waves}\label{sec:Polarization Properties}
The astrophysical or cosmological origins of stochastic background of GWs are usually assumed to be isotropic, unpolarized and stationary \citep{PhysRevD.46.5250,PhysRevD.59.102001,2000PhR...331..283M,PhysRevLett.116.061102,2019RPPh...82a6903C}. The GWs background at $z=0$ satisfy \citep{PhysRevD.59.102001}
\begin{equation}
  \langle h_{\lambda}(0,\omega)h_{\lambda}^*(0,\omega)\rangle = \delta(\omega-\omega')\delta^2(\hat{\Omega},\hat{\Omega}')\delta_{\lambda\lambda'}S_h(\omega),
  \label{GWsdensity}
\end{equation}
where $\delta(\Omega,\Omega')=\delta(\varphi-\varphi')\delta(\cos\theta-\cos\theta')$ is the covariant Dirac delta function on the two-sphere, and $S_h(\omega)$ is the spectral density of the stochastic background of GWs defined by
\begin{equation}
S_h(\omega) =  \frac{3H_0^2\Omega_\mathrm{gw}(\omega)}{\omega^3},
\end{equation}
where $\Omega_\mathrm{gw}$ is the fractional energy density spectrum of GWs, which is often assumed as a constant within radio bands in many stochastic background of GWs models, such as inflationary \citep{Starobinsky:1979ty,PhysRevD.50.1157}, cosmic string \citep{PhysRevLett.98.111101,2002PhLB..536..185S,1976JPhA....9.1387K,PhysRevD.71.063510}, pre-Big-Bang \citep{1995PhLB..361...45B,PhysRevD.55.3330,PhysRevD.73.063008} and PBH \citep{Maggiore:1999vm,PhysRevLett.102.161101}.
For unpolarized GWs, the mixed term $\langle h_+(0,\omega)h_\times^*(0,\omega) \rangle$ would vanishes while the square terms $\langle h_+^2(0,\omega)\rangle=\langle h_\times^2(0,\omega)\rangle\neq 0$. 

Cosmological GW sources, however, can generate polarized GWs by some mechanism, such as helical turbulence during a first-order phase transition \citep{PhysRevLett.95.151301,PhysRevD.92.043006,Ellis:2020uid,PhysRevD.102.083512,PhysRevResearch.3.013193}, gravitational chirality and modification \citep{PhysRevLett.101.141101,Alexander:2009tp}, pseudoscalar-like couplings between the inflation and curvature \citep{PhysRevLett.83.1506,PhysRevD.68.104012,Bartolo:2017szm} or gauge fields \citep{Sorbo:2011rz,PhysRevD.85.023525,Adshead:2013nka,Shiraishi:2013kxa,Dimastrogiovanni:2016fuu}. Then the spectral density $S(\omega)$ can be given by \citep{PhysRevLett.97.151101,galaxies10010034,2019MNRAS.487..562C}
\begin{eqnarray}
  S_h(\omega)&&=\begin{pmatrix}  \langle h_+(0,\omega)h_+^*(0,\omega)\rangle & \langle h_+(0,\omega)h_\times^*(0,\omega)\rangle  \\  \langle h_\times(0,\omega)h_+^*(0,\omega)\rangle & \langle h_\times(0,\omega)h_\times^*(0,\omega)\rangle \end{pmatrix}\nonumber\\
  &&=\begin{pmatrix}  I_g(\omega)+Q_g(\omega) & U_g(\omega)-iV_g(\omega)  \\  U_g(\omega)+iV_g(\omega) & I_g(\omega)-Q_g(\omega) \end{pmatrix},
\end{eqnarray}
where $I_g$, $Q_g$, $U_g$ and $V_g$ are the Stokes parameters of GWs defined as follows:
\begin{eqnarray}
  I_g&&=\langle h_+^2(0,\omega) \rangle+ \langle h_\times^2(0,\omega) \rangle,\nonumber\\
  Q_g&&=\langle h_+^2(0,\omega) \rangle- \langle h_\times^2(0,\omega) \rangle,\nonumber\\
  U_g&&=2\mathrm{Re}\langle h_+(0,\omega)h_\times^*(0,\omega) \rangle,\nonumber\\
  V_g&&=-2\mathrm{Im}\langle h_+(0,\omega)h_\times^*(0,\omega) \rangle.
  \label{StokesGWs}
\end{eqnarray}

The Stokes parameters of EMWs can be written as
\begin{eqnarray}
  I_\gamma&&=\sqrt{\frac{\epsilon_0}{\mu_0}}\left[\langle E_x^2(z,t) \rangle+ \langle E_y^2(z,t) \rangle\right],\nonumber\\
  Q_\gamma&&=\sqrt{\frac{\epsilon_0}{\mu_0}}\left[\langle E_x^2(z,t) \rangle- \langle E_y^2(z,t) \rangle\right],\nonumber\\
  U_\gamma&&=2\sqrt{\frac{\epsilon_0}{\mu_0}}\mathrm{Re}\langle E_x(z,t)E_y^*(z,t) \rangle,\nonumber\\
  V_\gamma&&=-2\sqrt{\frac{\epsilon_0}{\mu_0}}\mathrm{Im}\langle E_x(z,t)E_y^*(z,t) \rangle.
  \label{StokesEMWs}
\end{eqnarray}
Using the expressions of electric field $E_{x,y}(z,t)=-\partial_t A_{x,y}(z,t)$, the EMWs solutions in Eq. (\ref{EMWsolution}), and the Stokes parameters of GWs in Eq. (\ref{StokesGWs}), the EMWs Stokes parameters in Eq. (\ref{StokesEMWs}) become
\begin{subequations}
  \label{StokesEMWs2}
  \begin{eqnarray}
  I_\gamma&&=\sqrt{\frac{\epsilon_0}{\mu_0}} \int\dfrac{d\omega}{2\pi}\dfrac{l_{o}^2\omega^2}{4}\sin^2{\left(\dfrac{z}{l_{o}}\right)} (B_x^2+B_y^2)I_g,\label{StokesEMWs2I}\\
  %&&\approx\sqrt{\frac{\epsilon_0}{\mu_0}}\dfrac{cz\omega}{2\omega_{pl}^2}I_{g0}(B_x^2+B_y^2)\nonumber\\
  Q_\gamma&&=\sqrt{\frac{\epsilon_0}{\mu_0}} \int\dfrac{d\omega}{2\pi}\dfrac{l_{o}^2\omega^2}{4}\sin^2{\left(\dfrac{z}{l_{o}}\right)} \left[(B_y^2-B_x^2)Q_g-2B_xB_yU_g \right],  \label{StokesEMWs2Q}\\
  %&&\approx\sqrt{\frac{\epsilon_0}{\mu_0}}\dfrac{cz\omega}{2\omega_{pl}^2}[(B_y^2-B_x^2)Q_{g0}-2B_xB_yU_{g0}]\nonumber\\
  U_\gamma&&=\sqrt{\frac{\epsilon_0}{\mu_0}}\int\dfrac{d\omega}{2\pi}\dfrac{l_{o}^2\omega^2}{4}\sin^2{\left(\dfrac{z}{l_{o}}\right)} \left[(B_y^2-B_x^2)U_g+2B_xB_yQ_g \right],  \label{StokesEMWs2U}\\
  %&&\approx\sqrt{\frac{\epsilon_0}{\mu_0}}\dfrac{cz\omega}{2\omega_{pl}^2}[(B_y^2-B_x^2)U_{g0}+2B_xB_yQ_{g0}]\nonumber\\
  V_\gamma&&=\sqrt{\frac{\epsilon_0}{\mu_0}}\int\dfrac{d\omega}{2\pi}\dfrac{l_{o}^2\omega^2}{4}\sin^2{\left(\dfrac{z}{l_{o}}\right)} (B_x^2+B_y^2)V_g. \label{StokesEMWs2V}
  %&&\approx\sqrt{\frac{\epsilon_0}{\mu_0}}\dfrac{cz\omega}{2\omega_{pl}^2}V_{g0}(B_x^2+B_y^2).
\end{eqnarray}
\end{subequations}
The Stokes parameter $I_\gamma$ represents the total intensity of the EMWs, $Q_\gamma$ and $U_\gamma$ correspond to the linear polarization components, and $V_\gamma$ represents the circular polarization. The same definitions and interpretations apply to the Stokes parameters for GWs, neverthless, the Stokes $Q_\gamma$ and $U_\gamma$ transform as spin-2 quantities under rotation, whereas for GWs, the Stokes $Q_g$ and $U_g$ transform as spin-4 quantities \citep{galaxies10010034}.

The polarization states of EMWs can be modified by the magnetized plsama along the line of sight, a phenomenon known as the Faraday rotation effect. The effect can be quantified by Faraday depth \citep{1966MNRAS.133...67B,2005A&A...441.1217B}
\begin{equation}
  \phi(z)=\frac{e^3}{8\pi^2 \epsilon_0 m_e^2c^3}\int_z^L{n_eB_{z}}dl\approx k_\phi{n_eB_{z}}(L-z),
  \label{faradaydepth}
\end{equation}
where $k_\phi=\frac{e^3}{8\pi^2 \epsilon_0 m_e^2c^3}$. A positive Faraday depth implies that the parallel magnetic field is aligned with the propagation direction of the EMWs. For cosmological sources at redshift $z_s$, Eq. (\ref{faradaydepth}) can be modified to
\begin{equation}
  \phi(z_s)=k_\phi\int_{0}^{z_s}\frac{cn_e(z_s)B_z(z_s)dz_s}{H_0(1+z_s)^3\sqrt{\Omega_m(1+z_s)^3+\Omega_\Lambda}}.
\end{equation}
The observed polarization angle modified by Faraday effect, called Faraday rotation, can be expressed as
\begin{equation}
  \chi=\chi_0+\phi(z)\lambda^2,
  \label{polangle}
\end{equation}
where 
\begin{equation}
  \chi_0=\frac{1}{2}\arctan\frac{U_{\gamma}}{Q_{\gamma}}
\end{equation}
is the intrinsic polarization angle of EMWs. From Eq. (\ref{StokesEMWs2}) we can infer that the intrinsic polarization angle of EMWs is constant. 

Another observable quantity related to polarization of EMWS is the complex polarized intensity defined by $P(\lambda^2)=Q_\gamma(\lambda^2)+iU_\gamma(\lambda^2)=\Pi_L I_{\gamma0}e^{2i\chi}$, where $\lambda=2\pi c/\omega$ is the wavelength of EMWs, $\Pi_L=\sqrt{Q_{\gamma}^2+U_{\gamma}^2}/I_\gamma$ is the linear polarization degree of EMWs and $I_{\gamma0}$ is the normalized intensity given by 
\begin{equation}
  I_{\gamma0}=\int dz I_{\gamma}e^{2i\phi(z)\lambda^2}.
\end{equation}
The polarized intensity is subject to wavelength-dependent Faraday depolarization effects, which are governed by the properties of the intervening magnetized plasma. The depolarization can be quantified by factor \citep{1966MNRAS.133...67B,1998MNRAS.299..189S,2014A&A...567A..82S}
\begin{equation}
  \mathrm{DP}\equiv \frac{P(\lambda^2)}{P_0}=\frac{1-e^{2i\phi(z)\lambda^2-2\sigma_{\phi}^2\lambda^4}}{2i\phi(z)\lambda^2-2\sigma_{\phi}^2\lambda^4},
  \label{DP}
\end{equation}
where $P_0$ is the polarized intensity that has not been depolarization, and $\sigma_{\phi}^2$ is the dispersion of $\phi$ contributed by the spatial perturbations of electron density and magnetic field. Then the polarization intensity can be written as 
\begin{equation}
  P(\lambda^2)=\Pi_L I_{\gamma0}e^{2i\chi_0}\cdot\mathrm{DP}.\label{Polintensity}
\end{equation}

\begin{figure}[htb]
  \centering
  \includegraphics[width=0.8\textwidth]{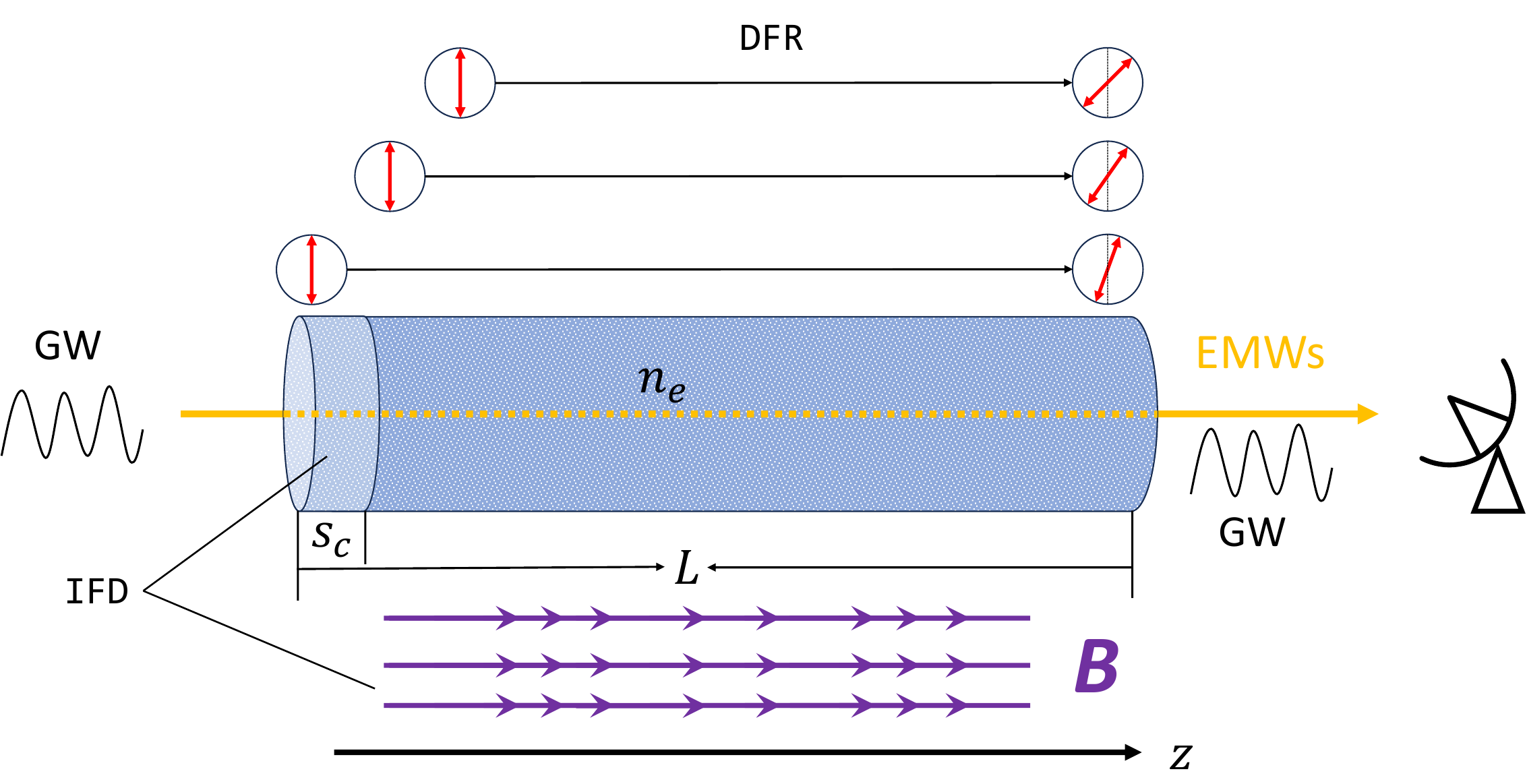}
  \caption{\label{fig:Depolarization} A simple sketch illustrating the concepts of DFR and IFD during the propagation of EMWs through a magnetized, perturbing plasma.  The vertical red vectors represent the initial phases at different locations, which undergo varying degrees of Faraday rotation as they propagate over different distances.  EMWs are generated within a small region $s_c$.}
\end{figure}

\begin{figure}[htb]
  \centering
  \includegraphics[width=0.8\textwidth]{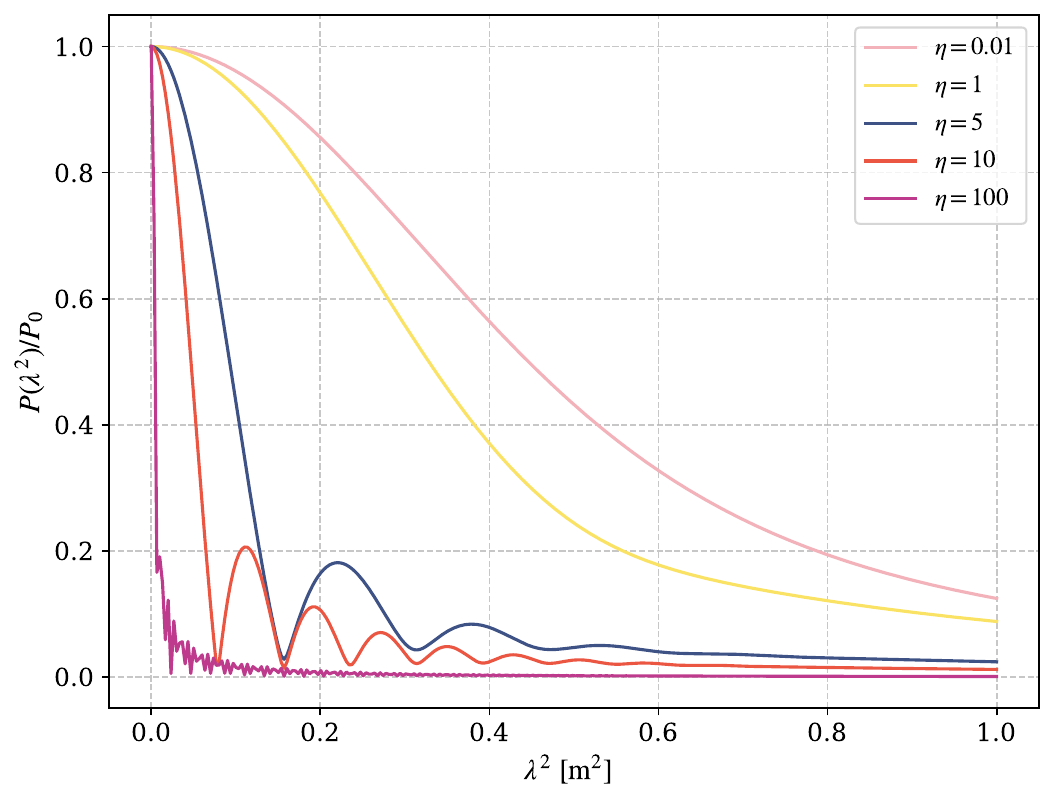}
  \caption{\label{fig:Polintensity} The depolarization effects of normalized polarized intensity with diverse $\eta$.}
\end{figure}

The average of Faraday depth is 
\begin{eqnarray}
  \langle \phi \rangle&&=k_{\phi}\int_{z}^{L} \langle n_e(z) B_z(z)\rangle dz\nonumber\\
  &&=k_{\phi}\int_{z}^{L}\langle [n_{e0}+\tilde{n}_e(z)] [B_{z0}+\tilde{B}_z(z)]\rangle dz\nonumber\\
  &&=k_{\phi}\int_{z}^{L}n_{e0}B_{z0}dz+k_{\phi}\int_{z}^{L}\langle \tilde{n}_{e}(z) \tilde{B}_{z}(z)\rangle dz\nonumber\\
  &&=k_{\phi}n_{e0}B_{z0}(L-z).
  \label{averagephi}
\end{eqnarray}
And $\sigma_{\phi}^2=\langle \phi^2\rangle-\langle \phi\rangle^2$, where $\langle \phi^2\rangle$ is
\begin{eqnarray}
  \langle \phi^2\rangle&& =k_{\phi}^2\iint _{z}^{L}dzdz'\langle n_e(z) B_z(z)n_e(z') B_z(z')\rangle \nonumber\\
  &&=k_{\phi}^2\iint _{z}^{L}dzdz' \langle n_{e0}^2B_{z0}^2+ \tilde{n}_{e}^2(z) \tilde{B}_{z}^2(z) \rangle\nonumber\\
  &&=k_{\phi}^2n_{e0}^2B_{z0}^2(L-z)^2+k_{\phi}^2\iint _{z}^{L}dzdz'\langle \tilde{n}_{e}^2(z)\rangle  \langle \tilde{B}_{z}^2(z)\rangle ,\nonumber\\
\end{eqnarray}
where $\langle \tilde{B}_{z}^2(z)\rangle$ and $\langle \tilde{n}_{e}^2(z)\rangle$ can be given by 
\begin{eqnarray}
  \langle \tilde{B}_{z}^2(z)\rangle &&= \frac{1}{3}\int d^3 qP_B(q_B),\nonumber\\
  \langle \tilde{n}_{e}^2(z)\rangle &&= \frac{1}{3}\int d^3 q P_{n_e}(q_n),
\end{eqnarray}
where $P_{n_e}(q_n)$ is the power spectrum of electron density fluctuation, which can also power-law spectrum \citep{1985ApJ...288..221C,1990ARA&A..28..561R}
\begin{equation}
  P_{n_e}(q_n)=C_n^2(s)q_n^{-\alpha_n}.
  \label{powerlawne}
\end{equation}
When electron density fluctuation and magnetic field fluctuation are driven by the same turbulent process, their wavenumbers and power-law spectral indices are consistent \citep{2003MNRAS.345..325C}, i.e. $q_n=q_B=q$ and $\alpha_n=\alpha_B=\alpha$. This is because the fluctuations are coupled through the turbulent velocity field, and their amplitudes scale proportionally with the velocity fluctuations, inheriting the same spectral index. Specifically, both power spectra usually follow the Kolmogorov scaling $P(q)\propto q^{-11/3}$, which has been support by observational evidence \citep{1981Natur.291..561A,2010ApJ...710..853C,Govoni:2017qmd} and numerical simulations \citep{PhysRevLett.88.245001,Beresnyak:2008ad}. Then $\sigma_{\phi}^2$ can be calculated by 
\begin{eqnarray}
  \sigma_{\phi}^2&&=k_{\phi}^2\iint _{z}^{L}dzdz'\langle \tilde{n}_{e}^2(z)\rangle  \langle \tilde{B}_{z}^2(z)\rangle \nonumber\\
  &&=k_{\phi}^2\iint _{z}^{L}dzdz'\int_{q_{\min}}^{q_{\max}}dqC_n^2 q^{2-\alpha}\int_{q_{\min}}^{q_{\max}}dqC_B^2q^{2-\alpha}\nonumber\\
  &&=\frac{9}{4}k_{\phi}^2C_n^2C_B^2(L-z)s_c\left(q_{\max}^{-2/3}-q_{\min}^{-2/3}\right)^2,\nonumber\\
  \label{sigmaphi}
\end{eqnarray}
where $q_{\min}=2\pi/L$ and $q_{\max}=2\pi/s_c$, with $s_c$ being the scale of the perturbation cell. 
A prominent mechanism, differential Faraday rotation (DFR), arises when polarized emission originates from multiple depths within a continuous medium. If the amplitudes of fluctuations are excessively smaller then the static terms, the imaginary term in Eq.(\ref{DP}) dominates, and DFR can be described by
\begin{equation}
  \mathrm{DP}_\mathrm{DFR}=\frac{\sin\left[2\overline{\phi} (z)\lambda^2\right]}{2\overline{\phi} (z)\lambda^2},
  \label{DPDFR}
\end{equation}
where $\overline{\phi}=\langle\phi \rangle $. 
In turbulent media, internal Faraday dispersion (IFD) dominates, generated by stochastic fluctuations in magnetic fields or electron density within the emitting region. The fluctuations terms are more dominated than the static terms and the real term in Eq.(\ref{DP}) dominates, The depolarization is quantified by
\begin{equation}
  \mathrm{DP}_\mathrm{IDF}=\frac{1-e^{-2\sigma_{\phi}^2\lambda^4}}{2\sigma_{\phi}^2\lambda^4}.
  \label{DPIFD}
\end{equation}
In contrast, external Faraday dispersion (EFD) operates in non-emitting foreground screens, where turbulent magnetic fields imprint an exponential suppression
\begin{equation}
  \mathrm{DP}_\mathrm{EDF}=e^{-2\sigma_{\phi}^2\lambda^4}.
  \label{DPEDF}
\end{equation}
When GWs propagate through magnetized plasma, the whole plasma cylinder can be regarded as an emitting region. EMWs generating from different position $z$ should experience DFR, and the perturbations of electron density and magnetic field should contribute to IFD (See Fig. \ref{fig:Depolarization}). 
Considering a region with total length $L$, containing $M=L/s_c$ perturbation cells, then the total polarization intensity in Eq. (\ref{Polintensity}) can be written into 
\begin{equation}
  P(\lambda^2)=  \Pi_L \langle I_{\gamma0}\rangle e^{2i\chi_{0}}\frac{1-e^{2i\overline{\phi} \lambda^2-2\sigma_{\phi}^2\lambda^4}}{2i\overline{\phi} \lambda^2-2\sigma_{\phi}^2\lambda^4},
  \label{Polintensity_tot}
\end{equation}
where
\begin{equation}
  \langle I_{\gamma0}\rangle=\sqrt{\frac{\epsilon_0}{\mu_0}} \int\dfrac{d\omega}{2\pi}\dfrac{\omega^2}{8}Ll_o (B_x^2+B_y^2)I_g(\omega).
\end{equation}
The polarization intensity exhibits  oscillatory behavior due to the imaginary exponent in Eq. (\ref{Polintensity_tot}) as $\lambda^2$ increases, while the the presence of the exponentially decaying term $e^{-2\sigma_{\phi}^2\lambda^4}$ rapidly suppresses these oscillations. Simultaneously, the overall magnitude of the expression decreases rapidly since the divisor grows proportionally to $\lambda^4$. As a result, the function approaches zero, with its decay following a $\lambda^{-4}$ scaling. Let $\eta=\overline{\phi} /\sigma_\phi^2$, the depolarization effect is illustrated in Fig. \ref{fig:Polintensity}. The total intensity $\langle I_{\gamma0}\rangle $ is assumed to be constant for simplicity. When $\eta$ is small, the depolarization effect exhibits a slow decay with mild oscillations, closely resembling the behavior described by Eq. (\ref{DPIFD}), indicating a weaker dependence on $\lambda^2$. As $\eta$ increases, the decay rate accelerates, and oscillations become more pronounced, gradually approaching the sinc-like form of Eq. (\ref{DPDFR}).

The alternative expression of $P(\lambda^2)$ can be written as a Fourier transform
\begin{equation}
  P(\lambda^2)=\int_{-\infty}^{+\infty} F(\phi) e^{2i\phi(z)\lambda^2}d\phi,
  \label{complexpol}
\end{equation}
where $F(\phi)$ is the Faraday dispersion function (FDF), which encodes the intrinsic polarized flux as a function of $\phi$. $F(\phi)$ can be obtained by inverse Fourier transform
\begin{eqnarray}
  F(\phi)&&=\int_{-\infty}^{+\infty} P(\lambda^2) e^{-2i\phi(z)\lambda^2}d\lambda^2.%\nonumber\\
  %&&=2\int_{0}^{+\infty} P(\lambda^2) e^{-2i\phi(z)\lambda^2}d\lambda^2,
  \label{F_phi}
\end{eqnarray}
The absolute value $\left\lvert F(\phi)\right\rvert $ denotes the polarization intensity of the EMWs. 
Eq. (\ref{F_phi}) provides a theoretical expression for FDF. However, in real observations, it is not always feasible to perfectly apply this expression because only a limited range of wavelengths is typically covered. In practice, the observed polarization intensity can be expressed as
\begin{equation}
  \widetilde{P} (\lambda^2)=W(\lambda^2)P(\lambda^2),
\end{equation}
where $W(\lambda^2)$ is the the observation window function. Specifically,  $W(\lambda^2)>0 $  when the wavelengths fall within the observable bands, and  $W(\lambda^2) = 0 $ otherwise.  Then the approximate FDF can be reconstructed by 
\begin{eqnarray}
  \widetilde{F} (\phi)&&=K\int_{-\infty}^{+\infty} \widetilde{P} (\lambda^2) e^{-2i\phi\lambda^2}d\lambda^2,%\nonumber\\
  %&&\approx K\sum_{m = 1}^{M} W_i(\lambda^2_m) P (\lambda_{m}^2) e^{-2i\phi\lambda_m^2},\nonumber\\
  \label{F_phi_approx}
\end{eqnarray}
where $K=\left(\int_{-\infty}^{+\infty} W(\lambda^2)d\lambda^2\right)^{-1}$ is the normalization for the observation window. 
\begin{figure}[htb]
  \centering
  \includegraphics[width=0.8\textwidth]{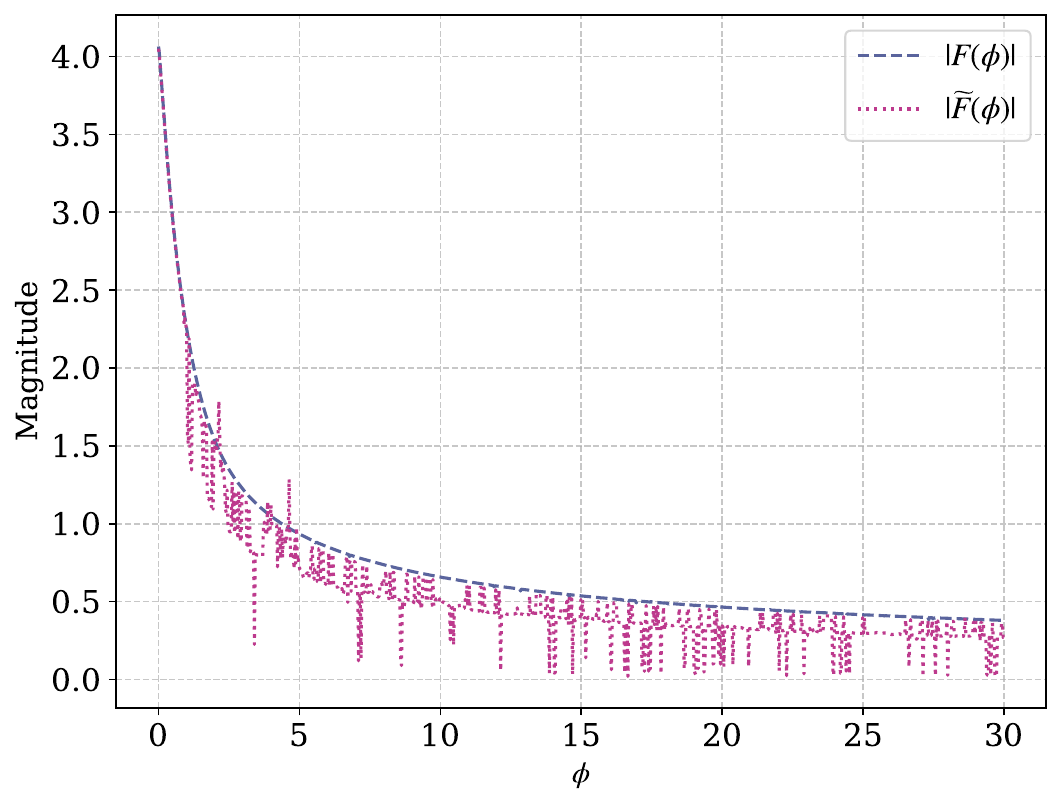}
  \caption{\label{fig:FDP_phi} Absolute value of the Faraday dispersion function in Eq (\ref{F_phi}) and Eq. (\ref{F_phi_approx}). The polarization intensity $P$, Faraday depth $\phi$, and $\sigma_{\phi}^2$ are normalized unit values.}
\end{figure}
Fig. \ref{fig:FDP_phi} is the absolute value of FDP defined by Eq (\ref{F_phi}) and Eq. (\ref{F_phi_approx}), respectively. The absolute value of FDP can be regarded as the polarization intensity contribution from different Faraday depth $\phi$. Due to the accumulative effect of GW-EMW over long propagation distance, EMW generated at small Faraday depth (longer propagation distances) undergo more significant Faraday rotation, resulting in a stronger contribution to the FDF. The oscillatory behavior at larger Faraday depth region is induced by the Faraday rotation factor  $e^{-2i \phi(z) \lambda^2}$.

Since the observation data are measured in discrete channels $\lambda_m^2\in [\lambda_{\min}^2,\lambda_{\max}^2]$, $m=1, 2, \cdots M$,  if the widths of squared wavelength $\delta\lambda^2\ll1/\phi$ for all channels, Eq. (\ref{F_phi_approx}) can be rewritten into\citep{1966MNRAS.133...67B,2005A&A...441.1217B,2009IAUS..259..591H,2012AJ....143...33A,galaxies6040140}
\begin{eqnarray}
  \widetilde{F} (\phi)&&\approx K_0\sum_{m = 1}^{M} W(\lambda^2_m) P (\lambda_{m}^2) e^{-2i\phi(\lambda_m^2-\lambda_0^2)},
  \label{F_phi_approx_discrete}
\end{eqnarray}
where 
\begin{eqnarray}
  K_0&&=\left[\sum_{m = 1}^{M} W(\lambda^2_m)\right]^{-1},\\
  \lambda_0^2&&=\frac{\sum_{m = 1}^{M} W(\lambda^2_m)\lambda_m^2}{\sum_{m = 1}^{M} W(\lambda^2_m)}.
\end{eqnarray}
Most of the telescopes have equal frequency channel bandwidths $\delta\nu$, which leads to the inequality of the channels width of squared wavelength $\delta\lambda^2$, and such differences can be ignored if $\delta\nu/\nu\ll1$.

The radio signals of the generated EMWs can be relatively weak, possibly just slightly above the level of noise, and we may mistake them as noise and dismiss them. Polarization can serve as a criterion to distinguish the EMWs generated via Gertsenshtein effect, as well as reveal the polarization state of the GWs. As we mentioned in the previous context, the generated EMWs can be accumulated by the within the perturbative electron densities and magnetic fields during the propagation. Longer propagation distances can result in greater rotation of the polarization plane, corresponding to larger Faraday rotation depth. The Faraday spectra of the signal should be similar to the shape illustrated in Fig \ref{fig:FDP_phi}. We can obtain the Stokes parameters and the polarization states of the EMWs by the polarization observation, while the Faraday depth and corresponding dispersions can be calculated by the model of electron density and magnetic field. The polarization angel $\chi_0$ and origin polarization intensity $P_0(\lambda^2)=\Pi_L I_{\gamma0}$ of the EMWs before being depolarized can be calculated by Eq. (\ref{polangle}) and Eq. (\ref{Polintensity}), respectively. Then, the origin linear components of EMWs are $Q_{\gamma0}=P_0\cos2\chi_0$ and $U_{\gamma0}=P_0\sin2\chi_0$. The Stokes parameters of GWs $(I_g, Q_g, U_g, V_g)$ can be revealed by the Stokes parameters of EMWs before depolarization $(I_{\gamma0},Q_{\gamma0},U_{\gamma0},V_{\gamma0})$ by Eq. (\ref{StokesEMWs2}). 

\section{Detection Sensitivity of the Telescopes}\label{sec:Sensitivity}
It is more generally to use the dimensionless characteristic strain amplitude $h_c(\omega)$ which is defined by \cite{Romano:2016dpx,2018JCAP...11..038K,PhysRevD.103.123541,2024ApJ...971L..10L}
\begin{equation}
  h_c(\omega)=\sqrt{\omega S_h(\omega)},
  \label{hc}
\end{equation}
where $S_h(\omega)$ here usually represents the total intensity $I_g(\omega)$ of GWs.  Using the constant energy density spectrum $\Omega_{\mathrm{GW}}$, Eq. (\ref{hc}) can be rewritten into
\begin{equation}
  h_c(\omega)=\sqrt{\frac{3H_0^2\Omega_{\mathrm{GW}}}{\omega^2}}.
  \label{OmegaGW_hc}
\end{equation} 
For broadband signals, the minimum detectable flux of a single-dish radio telescope can be given by \citep{2013tra..book.....W,2016era..book.....C,2017isra.book.....T}
\begin{equation}
  S_{\min}=\mathrm{SNR}_{\min}\frac{S_\mathrm{sys}}{\sqrt{n_\mathrm{pol}\Delta\nu \Delta t_\mathrm{obs}}},
  \label{Smin}
\end{equation}
where $\mathrm{SNR}_{\min}$ is the minimum signal-to-noise ratio, $n_\mathrm{pol}$ represents the number of polarization channels of the telescope, $\Delta\nu$ is the total frequency bandwidth, $\Delta t_\mathrm{obs}$ denotes the total observation time that the signals are detectable, and $S_\mathrm{sys}$ is the system equivalent flux density that can given by
\begin{equation}
  S_\mathrm{sys}=\frac{2k_BT_\mathrm{sys}(\omega)}{A_\mathrm{eff}},
  \label{SEFD}
\end{equation}
where $k_B$ is the Boltzmann constant, $A_\mathrm{eff}$ is the effective area of the radio telescope, and $T_\mathrm{sys}(\omega)$ is the frequency-dependent system temperature. For radio antenna array, the system equivalent flux density can be given by 
\begin{equation}
  S_\mathrm{sys}=\frac{2k_BT_\mathrm{sys}(\omega)}{\sqrt{n_a(n_a-1)}A_\mathrm{eff}},
\end{equation}
where $n_a$ is the number of antennas. $T_\mathrm{sys}(\omega)$ is the sum of all sources referenced to the input of a radiometer connected to the output of a radio telescope
\begin{equation}
  T_\mathrm{sys}(\omega)=T_\mathrm{sky}+T_\mathrm{atm}(\omega)+T_\mathrm{noise}(\omega),
\end{equation}
where $T_\mathrm{sky}$ is the continuum brightness temperature of the sky including CMB and non-thermal emission of MW and/or extragalactic background, $T_\mathrm{atm}(\omega)$ is the emission from the Earth's atmosphere, and $T_\mathrm{noise}(\omega)$ is the noise contribution from receiver. The MW contributions can be modeled by \citep{Platania:1997zn,deOliveira-Costa:2008cxd}
\begin{equation}
  T_\mathrm{MW}(\omega)=T_\mathrm{MW}(\omega_{*})\left(\frac{\omega}{\omega_*}\right)^{\beta_\mathrm{MW}},
\end{equation}
where $\omega_{*}=2\pi\times 408 \, \mathrm{MHz}$ is a referenced frequency \citep{1981A&A...100..209H,1982A&AS...47....1H}, and $\beta_\mathrm{MW}=-2.81$. While the extragalactic contribution can be modeled by \citep{2012ApJ...758...23C}
\begin{equation}
  T_\mathrm{ext}=T_\mathrm{ext,0}\left(\frac{\omega}{\omega_*'}\right)^{\beta_\mathrm{ext}},
\end{equation}
where $T_\mathrm{ext,0}=24.1 \, \mathrm{K}$, $\omega_*'=2\pi\times310 \, \mathrm{MHz}$ and $\beta_\mathrm{ext}=-2.599$. The temperature of the emission from the Earth's atmosphere is varied with the frequencies, and here we estimate $T_\mathrm{atm}(\omega)$ by the standard atmosphere model \cite{standard:Atmospheres,daidzic2015efficient} and obtain the temperatures using the \emph{am} program \cite{2022zndo...6774378P}. 
\begin{figure}[htp]
  \centering
  \includegraphics[width=0.8\textwidth]{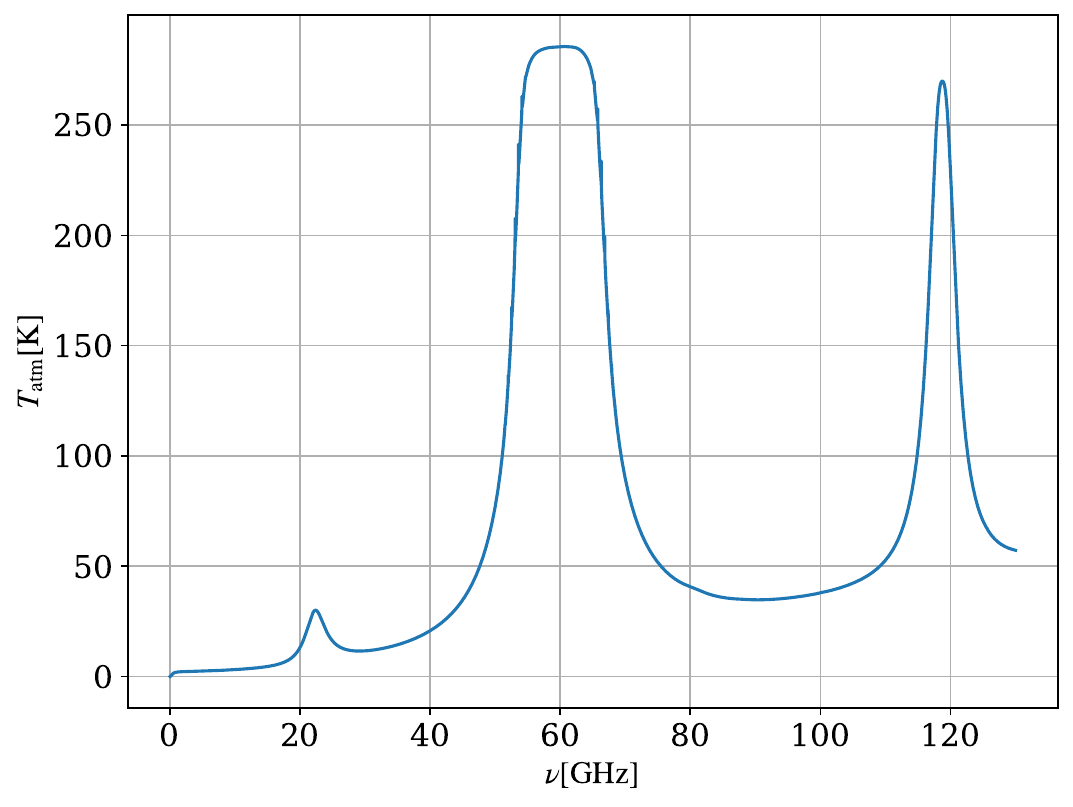}
  \caption{\label{fig:Tatm} Brightness temperatures of the emission from the Earth's atmosphere. The relative humidity is assumed to be 50\% in the troposphere.}
\end{figure}
Fig. \ref{fig:Tatm} illustrates $T_\mathrm{atm}(\omega)$ in different frequencies. The small peak at about 22 GHz is caused by the absorption of water-vapor in troposphere, while the high attenuation levels around 60 GHz and 118 GHz are caused by the oxygen absorptions \citep{long2015microwave}. The noise temperatures vary from different receivers of radio telescopes. Here we select three operating telescopes: The Five-hundred-meter Aperture Spherical Radio Telescope (FAST) \citep{2019SCPMA..6259502J,2020RAA....20...64J,2020Innov...100053Q}, Green Bank Telescope (GBT) \citep{GBTweb}, Parkes radio telescope (Parkes) \citep{Parkesweb}, and three next-generation radio interferometric arrays: FAST Core Array \citep{2024AstTI...1...84J}, Square Kilometre Array (SKA-mid) \citep{2019arXiv191212699B} and next-generation Very Large Array (ngVLA) \citep{2018ASPC..517...15S,2023AAS...24135702S}. 

The minimum detectable flux density of radio telescope $S_{\min}$ can be linked to the energy intensity $I_g(\omega)$ of the stochastic background of GWs and characteristic strain amplitude $h_c(\omega)$ by Eq. (\ref{StokesEMWs2I}) and (\ref{hc}). 
\begin{figure}[htb]
  \centering
  \includegraphics[width=0.8\textwidth]{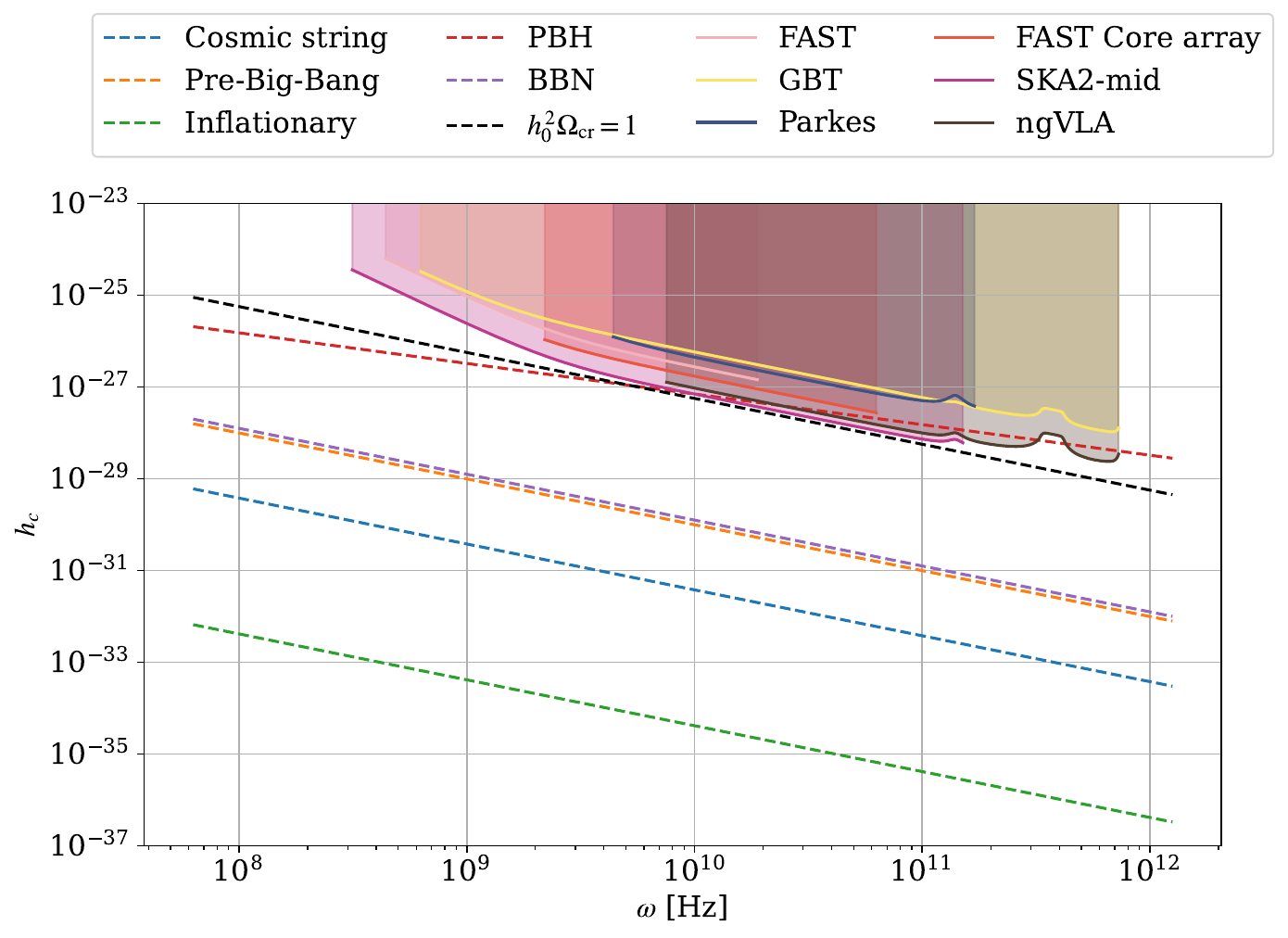}
  \caption{\label{fig:constrain} Upper Limits on the characteristic strain of the cosmological GW background from the $\sim$ 5-hour detection of GW-EMW conversion by telescopes and arrays. The colored solid lines indicate the detection limits of telescopes and arrays, and the dashed lines are some typical cosmological stochastic GW background sources and results, such as inflationary \citep[The GW tensor spectral index $n_T=0$]{Starobinsky:1979ty,PhysRevD.50.1157}, cosmic string \citep[The string tensor $G\mu=10^{-7}$ and the reconnection probability $p=0.1$]{PhysRevLett.98.111101,2002PhLB..536..185S,1976JPhA....9.1387K,PhysRevD.71.063510}, pre-Big-Bang \citep[The dilaton growth index in stringy phase $\mu=1.5$]{1995PhLB..361...45B,PhysRevD.55.3330,PhysRevD.73.063008}, PBH merger \citep[The mass of PBH $\sim10^{-4}M_{\odot}$] {2008MNRAS.390..192S,2015CQGra..32a5014M,2021LRR....24....4A}, and Big
Bang nucleosynthesis (BBN) \citep[The energy density $1.1\times10^{-5}$]{1997rggr.conf..373A,Maggiore:1999vm,2005APh....23..313C}. Here the $\mathrm{SNR_{\min}}$ is set to 1. The amplitudes and the spectral shapes of these GWs source can be remarkably diversified with different parameters.}
\end{figure}
Fig. \ref{fig:constrain} shows the detection limits of these observational equipments, together with some typical cosmological stochastic GW background sources. We can see that the detection sensitivities of SKA2-mid and ngVLA can reach the critical upper limit for stallar mass PBH within $\omega\sim 10^{10}-10^{11}$ Hz, since they have larger effective areas, enhanced gain, and larger frequency bandwidth coverages. These detection sensitivities have potential to make constrain on the mass of PBH, since the energy density spectrum of GWs generated by PBH is related to the mass. For example, the PBH merger model we use here can result in characteristic strain $h_c\propto M_\mathrm{PBH}^{5/6}\omega^{-2/3}$ \citep{2008MNRAS.390..192S,2015CQGra..32a5014M,2021LRR....24....4A}. We can also define the characteristic strain amplitudes of the two polarization modes of the GWs, $h_{c,+}$ and $h_{c,\times}$ by 
\begin{equation}
    h_c^2=h_{c,+}^2+h_{c,\times}^2,
\end{equation}
where $h_{c,+}$ and $h_{c,\times}$ can be represented by
\begin{eqnarray}
    h_{c,+}&&=h_c\sqrt{I_g+Q_g},  \\
    h_{c,\times}&&=h_c\sqrt{I_g-Q_g}. \label{hcpol}
\end{eqnarray}
As we obtain the Stokes parameters of EMWs before depolarization, the characteristic strain amplitudes in Eq. (\ref{hcpol}) can be linked to Eq. (\ref{StokesEMWs2I})-(\ref{StokesEMWs2U}). 
\section{Discussion and conclusion}\label{sec:DiscussionandConclusion}

In this work, we investigate the GW-EMW conversion process in turbulent magnetized plasma within MW and the expanding universe, with a focus on the accumulative conversion effects. For polarized GWs, we analyze the induced Faraday rotation and depolarization of the corresponding EMWs during propagation. Furthermore, we evaluate the detection sensitivity upper limits for both current operational telescopes and next-generation radio telescope arrays.

Our analytical framework can be extended to other EMW bands, contingent upon the availability of appropriate spectroscopic polarization observation capabilities. In the optical band, the Focal Reducer and Low Dispersion Spectrograph (FORS) instrument on the Very Large Telescope (VLT) achieves linear and circular polarization measurements with precision better than $10^{-3}$ \citep{1998Msngr..94....1A,2000SPIE.4008...96S,2020A&A...634A..70G}. In the X-ray regime, the Imaging X-ray Polarimetry Explorer (IXPE) provides simultaneous measurements of total intensity $I_\gamma$ and linear polarization components $(Q_\gamma , U_\gamma)$ within 2-8 keV energy range \citep{Bellazzini:2005ay,2021AJ....162..208S,2022JATIS...8b6002W}. The forthcoming enhanced X-ray Timing and Polarimetry (eXTP) mission will extend polarization measurements to 0.5-30 keV with systematic errors below 1\% \citep{eXTP:2016rzs}. 

At higher frequencies where photon wavelengths approach or exceed the electron Compton wavelength, classical phase-matching conditions for graviton-photon conversion may break down, necessitating QED corrections to the Gertsenshtein effect. The phase coherence is further influenced by the graviton mass. The phase matching condition can also affected by the mass of the graviton $m_g$, which modifies the group velocity as \citep{PhysRevD.85.043005,PhysRevD.88.064005,PhysRevD.89.104025}
\begin{equation}
  v_g(\omega)=c\sqrt{1-\frac{m_g^2c^4}{\hbar^2\omega^2}},
\end{equation}
where $\hbar$ is the reduced Planck constant. The mass of graviton introduces dispersion into Einstein's field equations Eq. (\ref{Einsteineq}) should be modified to Klein-Gordon equation \cite{Greiner:1997xwk}
\begin{equation}
  \left(\square - \frac{m_g^2c^2}{\hbar^2}\right)  \begin{pmatrix}  h_{+}  \\  h_{\times} \end{pmatrix}=\kappa^2 \begin{pmatrix} -B_x\partial_z A_y -B_y\partial_z A_x   \\ B_x\partial_z A_x -B_y\partial_z A_y \end{pmatrix}.
\end{equation}
This means smaller graviton mass and higher frequencies should have a better coherence response effect during the GW-EMW conversion. While current theories, experiments and observations have not yet been able to precisely determine the mass of the graviton, studies in these recent few years has already placed significant upper limits on its mass to levels as small as a few  $10^{-34}$ eV \citep{DeFelice:2021trp,DeFelice:2023bwq}, rendering dispersion effects negligible for our sensitivity estimates.

The perturbations of electron density and magnetic filed, which has been indicated by observational evidence and numerical simulations, also play critical role in the GW-EMW conversion efficiency, as well as the polarization features of EMWs response. In the interstellar medium of Milky Way, electron density fluctuation are relatively moderate, with typical fluctuations of $\widetilde{n}_e/n_{e0}\sim40\%$ \citep{2022A&A...668A..95W} in observation. Similarly, magnetic field fluctuations $\widetilde{B}  / B_0 $ within the Galactic ISM are $\gtrsim  30\%$ \citep{2017MNRAS.464L.105E}. Cosmological-scale intergalactic media exhibit even stronger inhomogeneities \citep{2006Sci...311..827I,2013A&ARv..21...62D}, which enhance conversion rates by disrupting GW-EMW coherence \citep{PhysRevLett.126.021104}. For polarized GW-EMW conversion, polarized EMWs can be depolarized by DFR and IFD due to the accumulation effect of GW-EMW conversion and the inhomogeneities in electron density and magnetic field. Since the frequencies of the telescopes/arrays we discuss above are relatively higher, the polarization intensity fo EMWs would not be reduced severely, especially at higher frequencies, where the polarization intensity is almost unaffected. The depolarization effect can be validated through multi-frequency campaigns using facilities like the Low-Frequency Array (LOFAR) \citep{2013A&A...556A...2V} and the upcoming SKA-low \citep{2019arXiv191212699B}. 

Finally, our analysis predicts the possible detection sensitivity for the upper limits of the cosmological GW sources. Although the constrain can not reach the BBN limit, which is currently regarded as the most sensitive bounds on the stochastic GW background, SKA2-mid and ngVLA still promise the potential to obtain constrain result that can be comparable to the levels of critical density $\Omega_\mathrm{cr}h^2=1$ and the PBH merger. It should be noticed that the estimation results can be also affected by the magnetic filed model and the redshift in the calculation. The co-moving magnetic field we use in this work is from \cite{2022MNRAS.515..256P} with a evolution parameter $\alpha$, and this parameter can be varied in the different simulation models. Additionally, the simulation results are only within $z\le2$, which is also the redshift we used in the estimation. We can anticipate the higher sensitivities of the next-generation telescope arrays to carry out observations at higher redshifts, thus obtain a more accurate cosmological magnetic field with higher redshifts.

\section*{ACKNOWLEDGMENTS}
We sincerely appreciate the referee’s suggestions, which helped us greatly improve our manuscript. This work was supported by  National SKA Program of China, No.2022SKA0110202 ,  National Key R\&D Program of China, No.2024YFA1611804 and China Manned Space Program through its Space Application System.


\bibliography{sample631}{}
\bibliographystyle{aasjournal}



\end{document}